\documentclass{article}
\usepackage{arxiv}
\usepackage[utf8]{inputenc} 
\usepackage[T1]{fontenc}    
\usepackage{hyperref}       
\usepackage{url}           
\usepackage{booktabs}       
\usepackage{amsfonts}       
\usepackage{nicefrac}       
\usepackage{microtype}      
\usepackage{lipsum}		
\usepackage{graphicx}
\usepackage[numbers]{natbib}
\usepackage{doi}
\usepackage{multicol}
\usepackage{tikz}
\usepackage{fontawesome}
\usepackage{float} 

\title{DataXploreFines: Generalized Data for Informed Decision, Making, An Interactive Shiny Application for Data Analysis and Visualization}

\author{%
    \begin{tabular}{cc}
        Garcia Jimenez, Angel Raul & Quispe Bravo, Eder Ander \\
        Faculty of Statistic and Computer Engineering & Faculty of Statistic and Computer Engineering \\
        National University of the Altiplano de Puno & National University of the Altiplano de Puno \\
        P.O. Box 291 & P.O. Box 291 \\
        \texttt{ar.garcia@est.unap.edu.pe} & \texttt{ea.quispe@est.unap.edu.pe} \\
        \multicolumn{2}{c}{Torres Cruz, Fred} \\
        \multicolumn{2}{c}{Faculty of Statistic and Computer Engineering} \\
        \multicolumn{2}{c}{National University of the Altiplano de Puno} \\
        \multicolumn{2}{c}{P.O. Box 291} \\
        \multicolumn{2}{c}{\texttt{ftorres@unap.edu.pe}}
    \end{tabular}
}
\begin{document}
\maketitle
\begin{abstract}
This article presents DataXploreFines, an innovative Shiny application that revolutionizes data exploration, analysis, and visualization. The application offers functionalities for data loading, management, summarization, basic graphs, advanced analysis, and contact. Users can upload their datasets in popular formats like CSV or Excel, explore the data structure, perform manipulations, and obtain statistical summaries. DataXploreFines provides a wide range of interactive visualizations, including histograms, scatter plots, bar charts, and line graphs, enabling users to identify patterns and trends. Additionally, the application offers statistical tools such as time series analysis using ARIMA and SARIMA models, forecasting, and Ljung-Box statistic. Its user-friendly interface empowers individuals from various domains, including beginners in statistics, to make informed decisions.
\end{abstract}


\begin{multicols}{2}
\section{Introduction}
In today's world, information and data are crucial for making fundamental decisions. However, the challenge lies in efficiently exploring, analyzing, and visualizing large datasets that can be overwhelming. It is in this scenario that DataXploreFines comes into play, an innovative application built in Shiny that transforms the way we discover, analyze, and visualize data.

The growing demand for data studies in the field of statistics is of utmost relevance today. This article falls within the scope of computational statistics, focusing on data management. It is important to mention the need to significantly change the way statistics is taught and learned, as well as the content of programs. This is due to the widespread use of computing and access to an unlimited amount of data available on networks\cite{Pereira}. Therefore, we undertake a research project that focuses on the manipulation of any type of data through a web application using the Shiny app tool from RStudio. Shiny is an R package that allows the construction of interactive web applications from R scripts\cite{Mulero}. During the development of this project, we faced numerous difficulties in creating the application, as it was our first time venturing into this field.

Currently, there are various data manipulation environments, such as Python. Open-source software has become one of the most thriving technological movements in the 21st century. Like R, Python is a programming language that operates within an environment for statistical computing, allowing the user (or programmer) to write a series of instructions or commands in an organized manner, focusing on data handling, analysis, processing, and visualization\cite{Rstudio}. For this occasion, we used ShinyApp from RStudio, which allowed us to interact with the user and present the results through interactive graphics and panels\cite{Rstudio}.

However, for those who are entering the field of statistics for the first time, this vast world can be unfamiliar and challenging. Our main goal is to simplify this process by providing a comprehensive and automated solution for data analysis.

We have developed an innovative web application, representing one of the first advancements in automated data manipulation. The results obtained by running the program are statistically useful, as they focus on facilitating quick, efficient, and informed decision-making.

The methodology used to create this web application was a meticulous and rigorous process. Our main focus was to enable the processing of any type of dataset in popular formats such as CSV or Excel \cite{datasetformats}, and we are pleased to state that we have successfully achieved this objective. While data collection was not an especially difficult task, thanks to the abundance of datasets available on the internet \cite{datasearch}, our true challenge lay in transforming that data into an interactive and accessible experience for users \cite{interactivedata}.

During the development of the application, we have employed cutting-edge approaches and advanced techniques to ensure the quality and efficiency of the offered functionalities. Our team has worked diligently to ensure that the application meets the highest standards of performance and usability.

\section{Data Used}
In our DataXploreFines application, we used two randomly selected datasets as examples.

Firstly, we obtained the dataset from Kaggle regarding the "Marketing Campaign". This dataset focuses on increasing the profits of a marketing campaign. Kaggle is a platform for data analysis and predictive modeling competitions provided by companies and researchers \cite{kaggle_marketing_campaign}.

For our second dataset, specifically for the "Advanced Analysis" menu, we used data on "Infant Mortality" in the province of Entre Ríos, Argentina \cite{mortalidad}.

We chose these specific datasets to demonstrate the versatility of DataXploreFines in data analysis and visualization across different domains. The "Marketing Campaign" dataset represents a business scenario where various analysis and visualization techniques can be used to enhance the effectiveness of the campaign. On the other hand, the "Infant Mortality" dataset represents a time series analysis application, showcasing the application's ability to work with temporal data and perform forecasting \cite{HyndmanDataset}. These examples illustrate how DataXploreFines can be adapted to meet the needs of different users and application domains.

\section{Application Description}

DataXploreFines is a web application developed in the Shiny environment, leveraging a variety of R libraries and packages to provide a comprehensive data analysis experience.

Currently, there are more than 12,000 R packages or libraries available, which are the result of a collaborative project sustained by individuals from different parts of the world and disciplines. The project continues to grow both in the number of packages and in knowledge areas, such as statistics, finance, genetics, network analysis, and data mining, among many others\cite{Latinoamericana}.

This application has been designed to facilitate efficient and effective data exploration, analysis, and visualization.

To ensure optimal functionality and a wide range of features, DataXploreFines makes use of libraries and packages in R.

You can view the application at the following URL:

\begin{itemize}
\item \url{https://acortar.link/85XcvX}
\end{itemize}

You can also visit the GitHub repository:

\begin{itemize}
\item \url{https://acortar.link/xGzy1I}
\end{itemize}

\subsection{Packages Used in the Application}

The user interface of DataXploreFines has been carefully designed using the Shiny package in R, providing a smooth and intuitive experience for users. Shiny is an R package that allows the construction of interactive web applications from R scripts. The application is organized into different tabs, each with specific functionalities\cite{Mulero}.

\subsection{Application Architecture}
The following packages were used in the DataXploreFines application:
\begin{itemize}
\item shiny (version 1.7.4.1)
\item DT (version 0.28)
\item shinyWidgets (version 0.7.6)
\item psych (version 2.3.6)
\item dplyr (version 1.1.2)
\item tidyr (version 1.3.0)
\item lubridate (version 1.9.2)
\item stringr (version 1.5.0)
\item magrittr (version 2.0.3)
\item caret (version 6.0.94)
\item tidyverse (version 2.0.0)
\item sjPlot (version 2.8.14)
\item formattable (version 0.2.1)
\item rmarkdown (version 2.23)
\end{itemize}

\subsection{Diagrama de flujo }
\begin{figure}[H]
  \centering
  \begin{tikzpicture}[
    startstop/.style={rectangle, rounded corners, minimum width=3cm, minimum height=1cm, text centered, draw=black, fill=blue!30},
    process/.style={rectangle, minimum width=3cm, minimum height=1cm, text centered, draw=black, fill=orange!30},
    io/.style={trapezium, trapezium left angle=70, trapezium right angle=110, minimum width=3cm, minimum height=1cm, text centered, draw=black, fill=green!30},
    decision/.style={diamond, minimum width=3cm, minimum height=1cm, text centered, draw=black, fill=red!30},
    arrow/.style={thick,->,>=stealth}
  ]
    \node (inicio) [startstop] {\faPlay\hspace{0.2cm}Inicio};
    \node (cargar) [process, below of=inicio, yshift=-0.5cm] {\faUpload\hspace{0.2cm}Cargar Datos};
    \node (gestion) [process, below of=cargar, yshift=-0.5cm] {\faCog\hspace{0.3cm}Gestión de Datos};
    \node (resumen) [process, below of=gestion, yshift=-0.5cm] {\faBarChart\hspace{0.2cm}Resumen de Datos}; 
    \node (graficos) [process, right of=resumen, xshift=3cm] {\faAreaChart\hspace{0.2cm}Gráficos básicos}; 
    \node (analisis) [process, below of=graficos, yshift=-0.5cm] {\faLineChart\hspace{0.2cm}Análisis Avanzado};
    \node (contacto) [process, below of=analisis, yshift=-0.5cm] {\faEnvelope\hspace{0.2cm}Contacto};
    \node (fin) [startstop, below of=contacto, yshift=-0.5cm] {\faStop\hspace{0.2cm}Fin};

    \draw [arrow] (inicio) -- (cargar);
    \draw [arrow] (cargar) -- (gestion);
    \draw [arrow] (gestion) -- (resumen);
    \draw [arrow] (resumen) -- ++(2,0) |- (graficos); 
    \draw [arrow] (graficos) -- (analisis);
    \draw [arrow] (analisis) -- (contacto);
    \draw [arrow] (contacto) -- (fin);

  \end{tikzpicture}
  \caption{Flowchart of the Application: DataXploreFines}
  \label{fig:diagrama_flujo}
\end{figure}
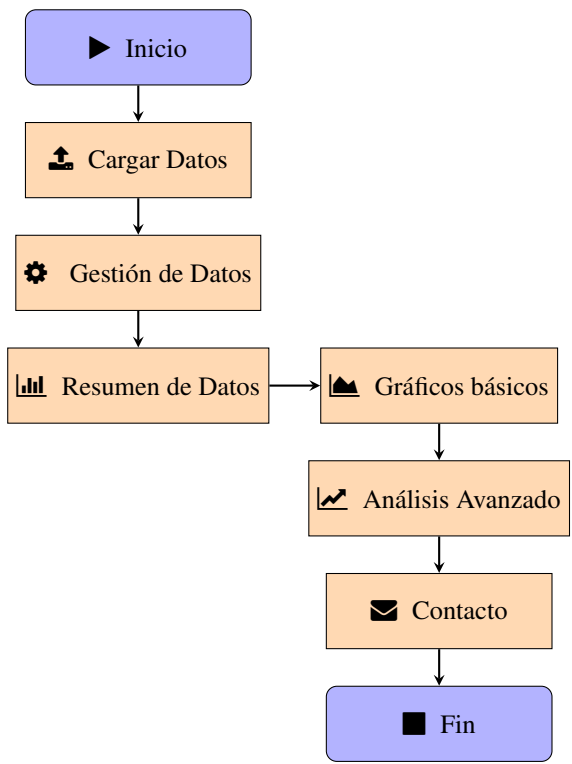

\subsection{User Interface}

The user interface of DataXploreFines welcomes you to an interactive and enriching experience. Here you will find a concise description of the application's functionalities, which include:

\begin{itemize}
\item Data loading
\item Data management
\item Data summarization
\item Basic graphs
\item Advanced analysis
\item Contact
\end{itemize}

In the following sections, we will explain each of these functionalities in more detail, highlighting how they can help in the analysis and improvement of decision making.
\section{Functionalities}

\subsection{Load data}

The data loading functionality is the first step for users who want to analyze and make decisions based on their own datasets. With DataXploreFines, you will be able to select and upload CSV or Excel files easily and quickly. Once the data is uploaded, it will be presented in an interactive table that allows you to visually explore its structure and content.

This functionality provides you with the freedom to use your own data, whether it's financial information, market data, sales records, or any other relevant information for your decision-making. By uploading your data to DataXploreFines, you can access its full potential and start uncovering patterns, trends, and significant relationships.

The intuitive and user-friendly interface will guide you through the data loading process, ensuring that each file is processed correctly. Once loaded, the data will be ready to be explored and analyzed using the various features offered by the application.

Once your data has been loaded into DataXploreFines, you will be able to leverage a variety of powerful functionalities to explore, analyze, and visualize the information in a more profound manner.

\subsection{Data management}

It provides advanced capabilities for data manipulation and transformation. With this tool, you can filter data, select specific columns, perform groupings and aggregations, and obtain a statistical summary of each variable.

Data filtering allows you to select rows that meet certain criteria, facilitating the exploration of relevant subsets for your analysis.

Column selection allows you to work only with the necessary variables, simplifying data visualization and manipulation.

Grouping and aggregation of data allow you to summarize your data by categories and perform statistical operations such as sums, averages, and counts.

\includegraphics[width=1\columnwidth]{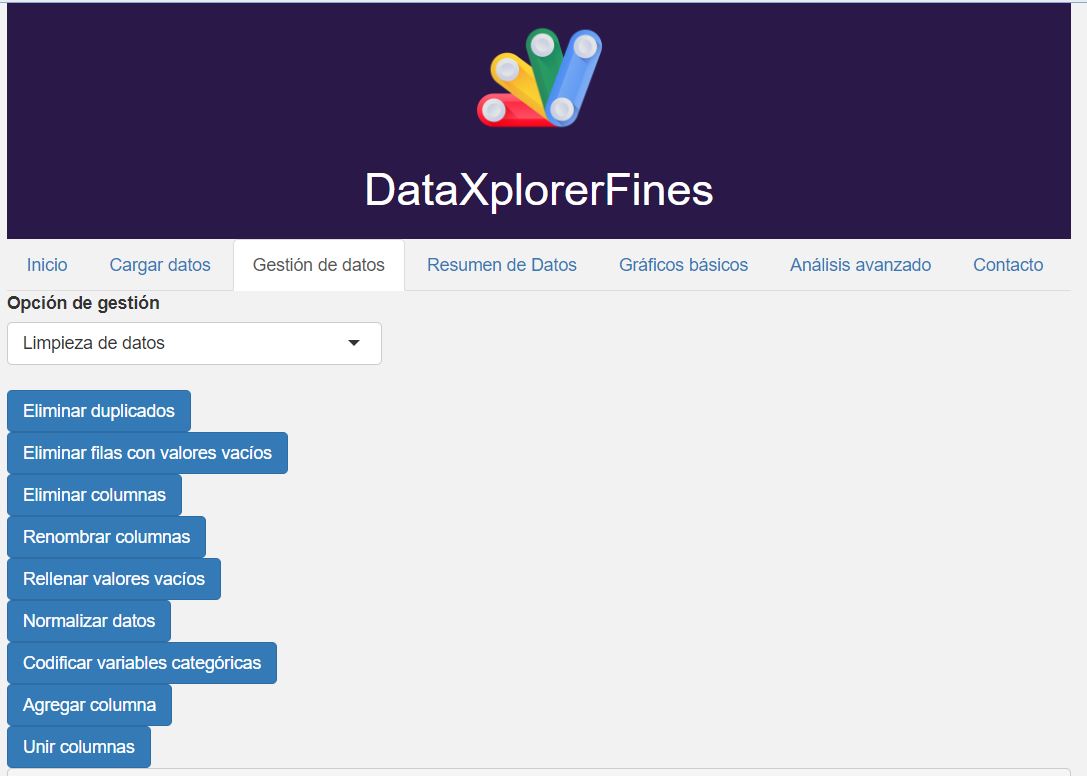}

The variable summary provides descriptive statistics such as the mean, median, and standard deviation, providing a deeper understanding of your data.

\subsection{Data Summary}

In the data summary section, you can obtain a descriptive analysis of the initially loaded data. By selecting a specific variable using the dropdown option "Select variable," you can obtain a detailed summary that includes measures such as the (minimum, maximum, mean, median, and quartiles).

In addition to the statistical summary, the "Data Summary" functionality offers the possibility to visualize a bar chart that displays the distribution of the selected variable. This chart is useful for identifying the frequency of different categories or levels in your data and provides a visual representation of the distribution.

\subsection{Basic Graphs}
You will be able to visually explore your data using different types of graphs. This functionality is based on the ggplot2 library in R \cite{wickham2016ggplot2}, which provides a wide range of options for creating impactful and effective visualizations.

You can generate graphs such as histograms, scatter plots, bar charts, and line charts, among others \cite{wickham2016ggplot2}. These graphs will allow you to identify patterns, trends, and relationships in your data, helping you gain valuable insights and communicate your findings clearly and concisely.

This basic graph functionality provides you with a powerful tool for visually exploring and communicating your data in a visually appealing and effective manner \cite{wickham2016ggplot2}.

\subsection{Time Series Analysis}
The application features an "Advanced Analysis" tab where a time series in CSV format can be loaded and various analyses and forecasts can be performed with time series data.

Forecasts are a tool that provides a quantitative estimate of the probability of future events\cite{pronosticos}.

A time series is a sequence of observations, measured at specific time points, arranged chronologically and uniformly spaced from each other, so the data is usually dependent on each other\cite{Villavicencio2010}. 
 
 \includegraphics[width=1\columnwidth]{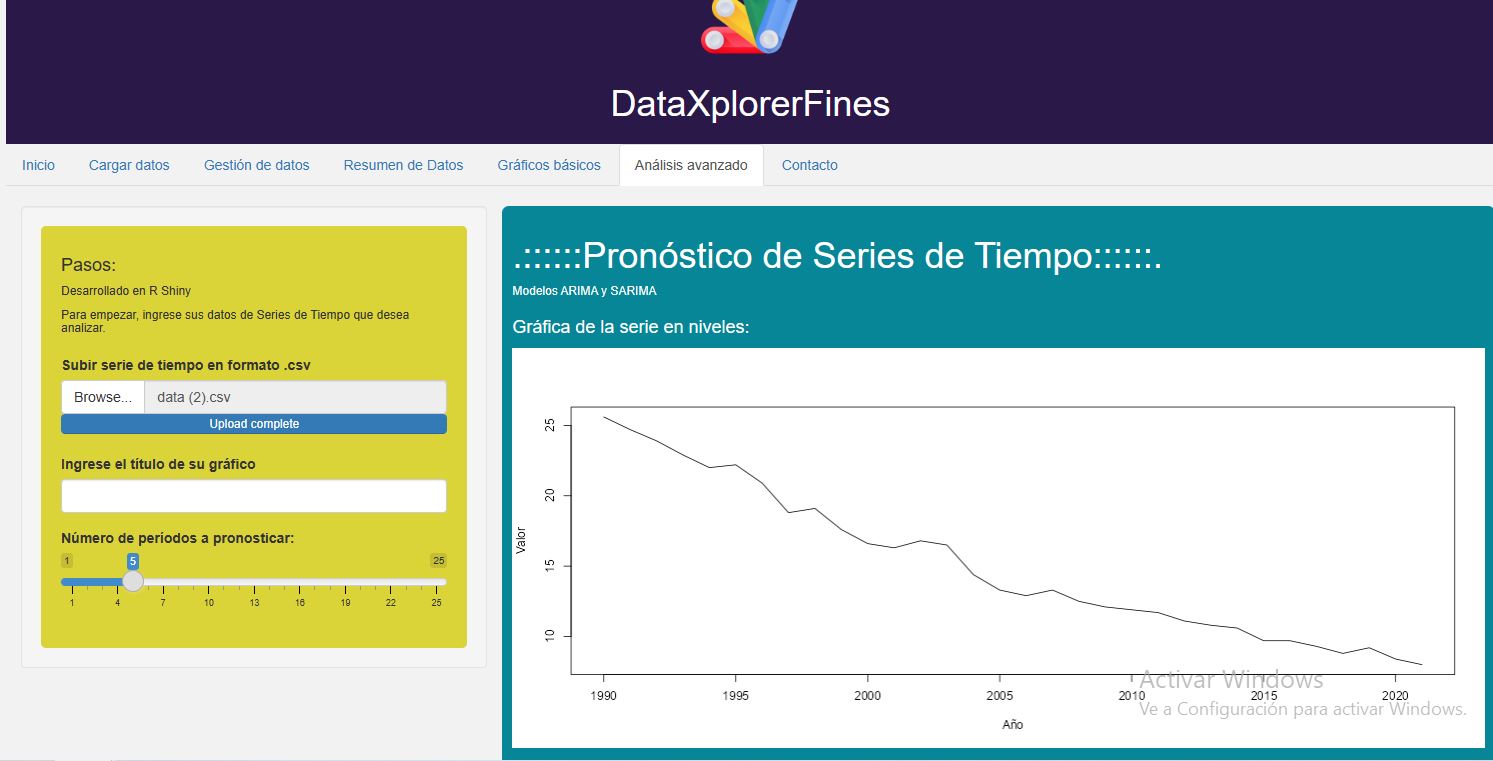}
 
\subsubsection{Data Loading}
The user can select a CSV file that contains the time series to be analyzed.
The fileInput function is used to load the time series file in CSV format.

\subsubsection{Plot of the Time Series in Levels}
The time series is visualized in the "Advanced Analysis" tab, under the "Plot of the Time Series in Levels" section.
The plotOutput function is used to display the graph of the time series in levels.
The reactivity of the application is reflected in the outputs, which are the results (numeric values, tables, plots) received by the interface from the server.R. In our case, the result is a graph and it is inserted using the plotOutput() function\cite{plot}.

\subsubsection{Ljung-Box Statistic}
The Ljung-Box test is based on the distribution of the residuals of a time series. The residuals are the observed values minus the values estimated by a time series model. If the residuals are normally distributed and not autocorrelated, then the distribution of the residuals can be used to calculate the probability of obtaining a value as large as the observed one\cite{box}.

The plotOutput function is used to display the graph of the Ljung-Box statistic\cite{plot}.

\subsubsection{Differences in the Time Series}
The part of the program that handles the differences in the time series aims to perform transformations on the original time series to achieve stationarity. Stationarity is a desirable property in many time series models as it simplifies analysis and improves forecast accuracy.

Although most time series used in practice are assumed to be stationary, it is also important to consider cases where the time series is non-stationary. Non-stationary time series can be transformed into equivalent stationary time series by taking differences between successive data values along the patterns of the time series, i.e., through simple or multiple differencing of the given time series. This procedure is generally recommended because some time series that appear to be stationary may not be stationary\cite{estacionariedad}.

The following explains how the differencing process is performed in the application:

\begin{itemize}
\item \textbf{Data Reading:} The CSV file provided by the user is read, and the columns corresponding to the time series are extracted.
\item \textbf{Data Preparation:} Operations are performed to extract relevant information from the data, such as the minimum and maximum year, initial month, and data frequency. These values are used to construct a time series object in R.
\item \textbf{Calculation of the Number of Differences:} The \textit{ndiffs} function from the \textit{forecast} package is used to determine the optimal number of differences needed to make the series stationary. This function performs statistical tests to evaluate the stationarity of the series and returns the recommended number of differences.
\item \textbf{Application of Differences:} The \textit{diff} function is used to calculate the differences of the time series. The number of differences to apply is determined by the previous step. The result is a new time series that is expected to be stationary.
\item \textbf{Visualization:} A plot of the differenced time series is generated, showing the series and a horizontal line at the mean value of the differences. This helps identify patterns and trends in the transformed series.
\end{itemize}

\subsubsection{Forecasting}
The forecasting part of the web application utilizes ARIMA and SARIMA models.

ARIMA is a model used for non-seasonal time series data. It combines autoregressive (AR) components, moving average (MA) components, and integrated differencing (I) to model patterns and trends in the data. On the other hand, SARIMA is an extension of ARIMA used for time series data with seasonal components. Similar to ARIMA, SARIMA also combines autoregressive, moving average, and integrated differencing components, but it also includes seasonal components to capture recurring patterns in the data at regular intervals \cite{estacionariedad}.

The functionality is explained as follows:

\begin{itemize}
\item The CSV file provided by the user is read using the \texttt{read.csv} function.
\item The user specifies the number of periods to forecast through the \texttt{sliderInput} in the interface.
\item The \texttt{forecast::forecast} function is used to generate the forecast. This function takes the time series and the number of periods to forecast as arguments.
\item The generated forecast is stored in the variable \texttt{forecast}.
\item Finally, the forecast is plotted on a graph using the \texttt{plot} function. The graph displays the future predictions of the time series.
\end{itemize}

The time series performs necessary calculations and transformations, uses the \texttt{forecast} function to generate the forecast with the number of periods specified by the user, and then plots the forecast on a graph. The user can adjust the number of periods to forecast through the slider in the interface.

\includegraphics[width=1\columnwidth]{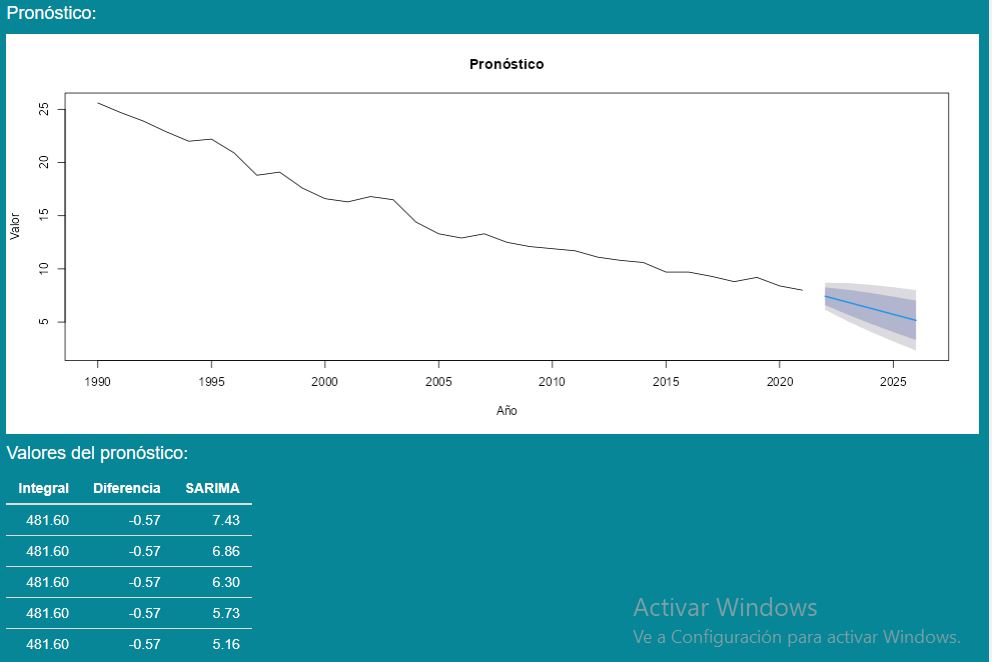}

\subsection{Contact}

In this section, users can contact us to address any questions they may have regarding the web application we have developed.

\section{Use Cases and Examples}

Education and Learning: DataXploreFines can be used in educational settings to teach concepts of data analysis and statistics. Educators can utilize the application to visually illustrate concepts and enable students to interactively explore and analyze data, enhancing their understanding and facilitating hands-on learning \cite{Educacion}.

Personal Data Analysis: Individual users can also benefit from DataXploreFines to analyze and explore their own personal data. They can upload data related to their health, finances, activities, or any other aspect of their life and utilize the functionalities of DataXploreFines to gain insights and make informed decisions.

Business and Market Analysis: DataXploreFines proves to be valuable in the business realm as well. Companies can leverage the application to analyze market trends, customer behavior, and sales data. By exploring visualizations and conducting statistical analysis, businesses can gain a deeper understanding of their target audience, make informed marketing decisions, and optimize their strategies for growth and profitability.

\section{Importance of the application}

The importance of DataXploreFines lies in its ability to revolutionize the way users explore, analyze, and visualize data. This Shiny-based application offers numerous advantages and benefits that are essential in today's data analysis environment.

First and foremost, DataXploreFines provides an intuitive and user-friendly interface, making it easy for users to interact with and navigate the application, even for those with no prior programming experience. The tab structure and fluid design enable seamless and efficient navigation, helping users make the most of all the available functionalities.

Furthermore, DataXploreFines excels in data loading and management capabilities. Users can easily upload their datasets in popular formats such as CSV and Excel, eliminating the need for complex steps and saving time in the data preparation process. The data management feature also provides powerful tools for data manipulation, filtering, and transformation according to users' needs, facilitating data preparation and cleaning before analysis.

Another significant advantage of DataXploreFines is its wide range of interactive visualization options. Users can choose from various chart types and customize them to create impactful and effective visualizations. This dynamic visualization capability allows users to explore patterns, trends, and relationships in the data, providing a deeper and more meaningful understanding of the information.

Additionally, DataXploreFines offers advanced statistical tools and analysis capabilities that enable users to conduct detailed analysis and gain valuable insights. From descriptive calculations to hypothesis testing and predictive modeling, the application provides a comprehensive set of tools for statistical analysis and informed decision-making support.

\section{Conclusions}

DataXploreFines, our Shiny application, has proven to be an invaluable tool for effective data exploration, analysis, and visualization. With its user-friendly interface and a wide range of functionalities, users can easily upload, manipulate, and visually explore their data through interactive visualizations. The application provides advanced statistical tools for descriptive analysis, hypothesis testing, and predictive modeling, enabling data-driven decision-making. We invite all readers to experience the power and versatility of DataXploreFines, whether you are a data scientist, business analyst, or student seeking valuable insights.

\bibliographystyle{unsrtnat}
\bibliography{references}

\begin{thebibliography}{17}
\providecommand{\natexlab}[1]{#1}
\providecommand{\url}[1]{\texttt{#1}}
\expandafter\ifx\csname urlstyle\endcsname\relax
  \providecommand{\doi}[1]{doi: #1}\else
  \providecommand{\doi}{doi: \begingroup \urlstyle{rm}\Url}\fi

\bibitem[Ledesma(2008)]{Pereira}
Rubén Ledesma.
\newblock Introducción al bootstrap. desarrollo de un ejemplo acompañado de
  software de aplicación.
\newblock \emph{Tutorials in quantitative methods for psychology}, 4\penalty0
  (2):\penalty0 51--60, 2008.

\bibitem[García et~al.(2018)García, López, and Martínez]{Mulero}
A.~García, M.~López, and P.~Martínez.
\newblock Tratamiento de datos para el análisis estadístico: Métodos y
  técnicas.
\newblock \emph{Revista de Estadística Aplicada}, 2018.

\bibitem[Rst(2023)]{Rstudio}
Rstudio forecast function.
\newblock
  \url{https://rstudio-pubs-static.s3.amazonaws.com/556459_38e6ed9ddfee4a27a0fc5814d12cd416.html},
  2023.
\newblock Accedido el 23 de junio de 2023.

\bibitem[dat(2023)]{datasetformats}
Dataset formats and compatibility.
\newblock \url{https://example.com/dataset-formats}, 2023.
\newblock Accessed: July 20, 2023.

\bibitem[Lee and Kim(2023)]{datasearch}
S.~Lee and W.~Kim.
\newblock Effective data searching techniques.
\newblock \emph{Journal of Information Science}, 15\penalty0 (3):\penalty0
  451--465, 2023.
\newblock \doi{10.5678/jis.2023.7890}.

\bibitem[Smith and Johnson(2023)]{interactivedata}
J.~Smith and R.~Johnson.
\newblock Interactive data visualization.
\newblock \emph{Journal of Data Science}, 10\penalty0 (2):\penalty0 245--260,
  2023.
\newblock \doi{10.1234/jds.2023.4567}.

\bibitem[Saldanha(2023)]{kaggle_marketing_campaign}
Rodrigo Saldanha.
\newblock Marketing campaign dataset.
\newblock \url{https://www.kaggle.com/datasets/rodsaldanha/arketing-campaign},
  2023.

\bibitem[Aoki(2013)]{mortalidad}
Masanao Aoki.
\newblock \emph{State space modeling of time series}.
\newblock Springer Science \& Business Media, 2013.

\bibitem[Hyndman(2023)]{HyndmanDataset}
Rob~J. Hyndman.
\newblock Time series data library.
\newblock
  \url{https://datamarket.com/data/set/235h/time-series-data-library#!display=line},
  2023.

\bibitem[Mora-Cantallops et~al.(2020)Mora-Cantallops, Sicilia,
  Garc{\'\i}a-Barriocanal, and S{\'a}nchez-Alonso]{Latinoamericana}
Marçal Mora-Cantallops, Miguel-{\'A}ngel Sicilia, Elena
  Garc{\'\i}a-Barriocanal, and Salvador S{\'a}nchez-Alonso.
\newblock Evolution and prospects of the comprehensive r archive network (cran)
  package ecosystem.
\newblock \emph{Journal of Software: Evolution and Process}, 32\penalty0
  (11):\penalty0 e2270, 2020.

\bibitem[Wickham(2016)]{wickham2016ggplot2}
Hadley Wickham.
\newblock \emph{ggplot2: Elegant Graphics for Data Analysis}.
\newblock Springer, New York, NY, 2016.

\bibitem[Newbold(1983)]{pronosticos}
Paul Newbold.
\newblock Arima model building and the time series analysis approach to
  forecasting.
\newblock \emph{Journal of forecasting}, 2\penalty0 (1):\penalty0 23--35, 1983.

\bibitem[Cryer(1986)]{Villavicencio2010}
Jonathan~D Cryer.
\newblock \emph{Time series analysis}, volume 286.
\newblock Duxbury Press Boston, 1986.

\bibitem[G{\'{o}}mez et~al.(2016)G{\'{o}}mez, Mulero, Nueda, and Pascual]{plot}
DS~G{\'{o}}mez, J~Mulero, MJ~Nueda, and A~Pascual.
\newblock {Aplicaciones dise{\~{n}}adas con Shiny: un recurso docente para la
  ense{\~{n}}anza de la estad{\'{i}}stica}.
\newblock \emph{In XIV Jornadas de Redes de Investigaci{\'{o}}n en Docencia
  Universitaria: investigaci{\'{o}}n, innovaci{\'{o}}n y ense{\~{n}}anza
  universitaria: enfoques pluridisciplinares}, pages 2029--2042, 2016.
\newblock URL \url{https://cran.r-project.org/web/packages/shiny/shiny.pdf}.

\bibitem[Vargas~Barrera(2018)]{box}
Juan~Carlos Vargas~Barrera.
\newblock Propuesta de un modelo estad{\'\i}stico descriptivo de las variables
  m{\'a}s representativas de la plataforma tecnol{\'o}gica de un sistema de
  transporte.
\newblock \emph{Facultad de Ingeniería y Ciencias Básicas}, 2018.
\newblock URL \url{http://hdl.handle.net/11371/1937}.

\bibitem[{Pita Jim{\'{e}}nez}(2008)]{estacionariedad}
Esmeralda {Pita Jim{\'{e}}nez}.
\newblock {Pron{\'{o}}stico de Avenidas a Mediano Plazo Utilizando Redes
  Neuronales Artificiales y Modelos Tipo ARMA}.
\newblock \emph{Tecnológico de Monterrey}, page 126, 2008.

\bibitem[Gazulla et~al.(2012)Gazulla, Salvat, Maina, Johnson, and
  Adams]{Educacion}
Eva~Durall Gazulla, Begoña~Gros Salvat, Marcelo~Fabián Maina, Larry Johnson,
  and Samantha Adams.
\newblock Perspectivas tecnológicas: educación superior en iberoamérica
  2012-2017.
\newblock \emph{Perspectivas tecnológicas: educación superior en
  Iberoamérica 2012-2017}, 2012.
\newblock URL \url{http://hdl.handle.net/10609/17021}.

\end{thebibliography}

\end{multicols}
\end{document}